# Application of Chaotic Number Generators in Econophysics

Carmen Pellicer-Lostao[1], Ricardo López-Ruiz[2]

*Department of Computer Science and BIFI, Universidad de Zaragoza, 50009 - Zaragoza, Spain.*

*e-mail address:* [carmen.pellicer@unizar.es](mailto:carmen.pellicer@unizar.es)[1] , [rilopez@unizar.es](mailto:rilopez@unizar.es)[2] ; **Lecture Notes.**

**ABSTRACT:**

Agent-based models have demonstrated their power and flexibility in Econophysics. However their major challenge is still to devise more realistic simulation scenarios. The complexity of Economy makes appealing the idea of introducing chaotic number generators as simulation engines in these models. Chaos based number generators are easy to use and highly configurable. This makes them just perfect for this application.

## 1. INTRODUCTION

Economy is difficult to model. Intermittent crisis have demonstrated throughout decades, the inherent complexity of economic systems and the fatal flaws of contemporary models. Nevertheless, significant methodologies have constantly been developed to move forward. In today's Economy there are two dominant paradigms; one is based on Econometry and the other on dynamic stochastic general equilibrium [1]. But Economy seems unpredictable and quantitative econometric predictions fail not far in time. Also equilibrium based models have shown their limitations to predict disruptive crisis of real economies.

To this particular matter, Econophysics offer agent based models. These provide computerized simulation scenarios of great number of decision makers (agents) and institutions, which interact through prescribed rules. Computational power gives to these models a richness of scenarios not necessarily in equilibrium, able of reproducing more realistic and complex situations and handling a far wider range of non linear behaviour [2]. However, agent based models are not a panacea. Their major challenge lies on specifying how agents behave and in choosing the rules they use to make decisions. In fact, simplifying these specifications may lead to impractical simulation scenarios.

One essential constituent of many agent based models is their stochastic nature, for one way or another they may use stochastic simulation tools to run [3,4]. Typically a number of agents are randomly selected or awakened from sleep to take economic decisions. Not only agents, but also the economic variables they possess may be changed randomly. The random approach can be very handy as a simulation engine and it also takes into account the variety and unpredictability ingredients of real situations.

However, random number generators provide a statistical uniformity that may unwillingly disguise real situations. Going a little bit further, let's say that Economy may be complex but it hardly seems random. To contribute with a new approach, this work illustrates how chaos based number generators can be used in agent based models. As it will be seen, these tools are highly powerful and flexible. They offer the possibility of producing numbers with different statistical random quality or even chaotic. Moreover, they possess a series of control parameters to manipulate the dynamics of the simulation.

This paper explores the main characteristics of chaos based number generators and gives an illustrative example of their application in Econophysics dedicated to the ideal gas-like models for wealth distributions.

## 2. CHAOTIC BASE NUMBER GENERATORS

In general sense, the term chaos refers to physical phenomena that are fully deterministic and even so, unpredictable and erratic [5]. Determinism and unpredictability, the two essential components of chaotic systems seem to be present in real economic systems. On one hand, in the short term or at micro level, economic transactions are mainly deterministic, based on 'rational expectations' to maximize the long-run personal advantage. On the other, in the long term or at the macro level, economic variables seem to be capricious and erratic. This makes chaos an appealing feature to be considered in economic models.

Since 1990 many pseudo-random number generators based in chaotic dynamical systems have been proposed [6]. They are based on N-dimensional deterministic discrete-time dynamical systems, and this makes them able to offer a rich variety of possibilities as simulation engines in Monte Carlo processes. An N-dimensional deterministic discrete-time dynamical system is an iterative map *F: $R^N \rightarrow R^N$* of the form:

$$X_{k+1} = F(X_k, \Lambda) \qquad (1)$$

where *k = 0, 1,…,n* is the discrete time, *$X_0, X_1,…,X_n$*, are the states of the system at different instants of time and *Λ* is a vector of control parameters. This kind of systems present different asymptotic behaviours for different values of *$X_0$* and *Λ*, where nearby orbits converge to given compact sets of the space state called attractors. If these attractors display complex behaviour, they are said to be strange attractors. Moreover when the system exhibits sensibility to initial conditions (*$X_0$*) is said to be in chaotic regime.

To build a chaotic PRNG is necessary to construct a numerical algorithm that transforms the states of the system in chaotic regime into integer numbers, or typically bits. The existing designs use different techniques to pass from the continuum to the binary world:

1. Extracting bits from each state along the chaotic orbits.
2. Dividing the phase space into m sub-spaces, and output a number *i = 0, 1,…,m* if the chaotic orbit visits the *i-th* subspace.
3. Combining the outputs of two or more chaotic systems to generate the pseudo-random numbers.

The construction of a chaos based PRNG involves a series of design parameters (such as for example, the number of bits extracted with technique 1). These design parameters are able to control the statistical quality of randomness of the numbers produced. These generators can also produce chaotic numbers, extracted directly at the output of the dynamical system. The dynamical properties of these chaotic numbers can be controlled by $\Lambda$, vector of control parameters. As a result, chaotic based number generators constitute a straightforward tool for computational simulation experiments.

## 3. EXPERIMENTS IN ECONOPHYSICS OF WEALTH DISTRIBUTIONS

This section shows an example that applies a chaos based number generator to the traditional gas-like model of wealth distributions. In this model [3], a community of $N$ agents is given an initial equal quantity of money, $m_0$ to each agent. The total amount of money, $M=N*m_0$, is conserved. The system evolves a total time of $T\sim N^2$ transactions to obtain equilibrium. For each transaction, at a given instant $t$, a pair of agents $(i,j)$ with money $(m^t_i, m^t_j)$ is selected randomly and a random amount of money $\Delta m$ is traded between them.

In our simulation scenario $N=256$ agents, $m_0= 1000\$$ and $T=150000$ interactions will be taken. Additionally, a particular rule of trade is selected for its simplicity and extensive use [3]. This rule establishes that for each interaction a random the quantity of money $\Delta m$ is traded between agents $x_t$ and $y_t$ through the following equations:

$$\Delta m = \mu (m^t_i + m^t_j)/2 \qquad m^{t+1}_i = m^t_i - \Delta m \qquad m^{t+1}_j = m^t_j + \Delta m \qquad (2)$$

where $\mu$ is random number in the interval $[0,1]$ from a standard random generator. Here, the transaction of money is quite asymmetric as for a transaction, agent $j$ is the absolute winner, while $i$ loses a fraction of his money. Also, if agent $i$ has not enough money $(m^t_i < \Delta m)$ no transfer takes place.

As a new approach, a chaotic based number generator is going to be used to produce pairs of numbers to select pairs agents $(i, j)$ for every interaction. The generator selected is proposed in [6] and its functional structure is shown in Fig.1.

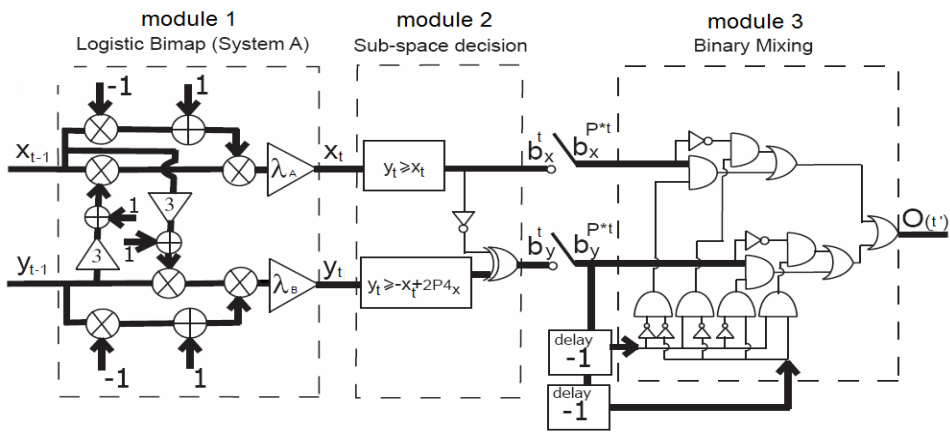

**Figure 1. Functional structure of the chaotic number generator used in the experiment.**



This chaos based number generator is built in three modules:

1. The chaotic module, where a **2D** chaotic map of logistic type produces real **2D** numbers $X_t=[x_t, y_t]$ in the interval **[0,1]** at any instant **t** of time.
2. The sub-space decision module. Here the phase space is divided in **4** sub-spaces and two binary numbers are produced $[b^t_X, b^t_Y]$ when the chaotic orbit visits the any of the **4** subspaces. The binary numbers are sampled with a factor **P** to control the correlation of consecutive outputs.
3. The binary mixing module. It combines the two bits obtained in the previous module to produce the output bit **O(t')**. This block performs a binary operation that allows good statistical random quality in the output sequence.

As it can be observed, this generator has three parameters for control. On one hand, there is a design parameter, the sampling factor **P**, that controls the statistical quality of the random output bits. On the other, the chaotic map has two control parameters $\lambda_A$ and $\lambda_B$ that can vary as desired the chaotic attractor and produced different orbits, and so different patterns of chaotic numbers.

For simulation proposes, at every instant **t** a pair of agents **(i, j)** will be selected for interaction using two integer numbers in the range **1** to **256** produced by the generator. The generator is able to produce **8** bits numbers from the binary sequence **O(t')** with different statistical random quality depending on the sampling factor **P** selected [6]. The generator is also able to generate chaotic numbers when the output is taken directly from the chaotic block (module 1) with a simple float to integer conversion:

$$i = (int)(x_t * N)+1 \qquad j = (int)(y_t * N)+1 \qquad (3)$$

Fig. 2 shows how the generator is able to produce integer numbers with different statistical quality of randomness depending on the sampling factor **P**. The figure also shows how the generator can produce different chaotic patterns of integers depending on the control parameters $\lambda_A$ and $\lambda_B$.

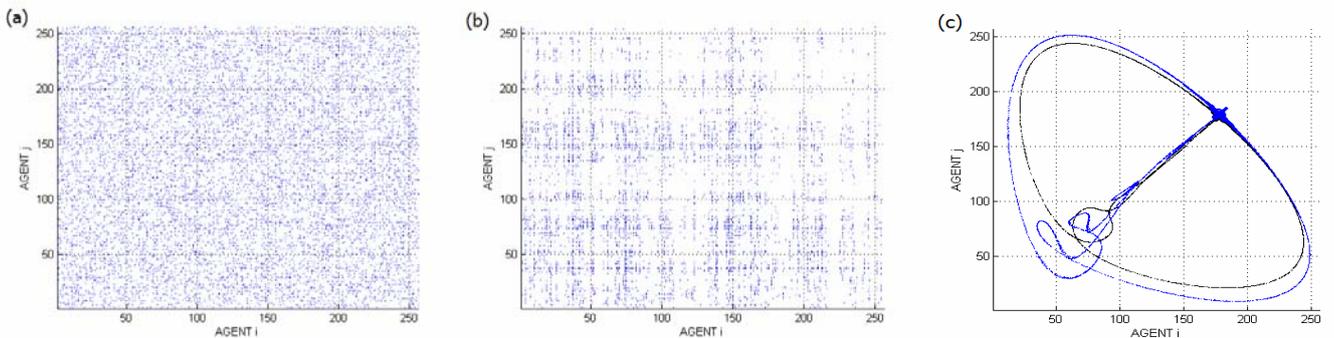

**Figure 2. Ten thousand pairs of one byte integers produced the chaotic number generator (a) random integers produced with P=45. (b) random integers of poor statistical quality with P=3. (c) two sets of chaotic integers produced with two sets of control parameters $\lambda_A=\lambda_B=1.05032$ (black line) and $\lambda_A=1.05032$ $\lambda_B = 1.08492$ (blue line).**

Three scenarios are considered to highlight the power of the generator for simulating different market interactions. A further study of different scenarios produced with this generator can view viewed in [7]. As the random selection of agents has already been tested [3], experiments will be done with a selection of agents of low random statistical properties (the sampling factor will be set to *P=3*). Also, two sets of chaotic numbers with two different sets of control parameters will be considered, to modify the symmetry of the chaotic attractor *($\lambda_A=\lambda_B=1.05032$* (symmetric) *$\lambda_A=1.05032$ $\lambda_B = 1.08492$* (asymmetric)). As it can be seen in Fig. 2 (c), the symmetry of the attractor controls the complex network of economic interactions in the community.

When simulations are run, the results are shown in Fig. 3:

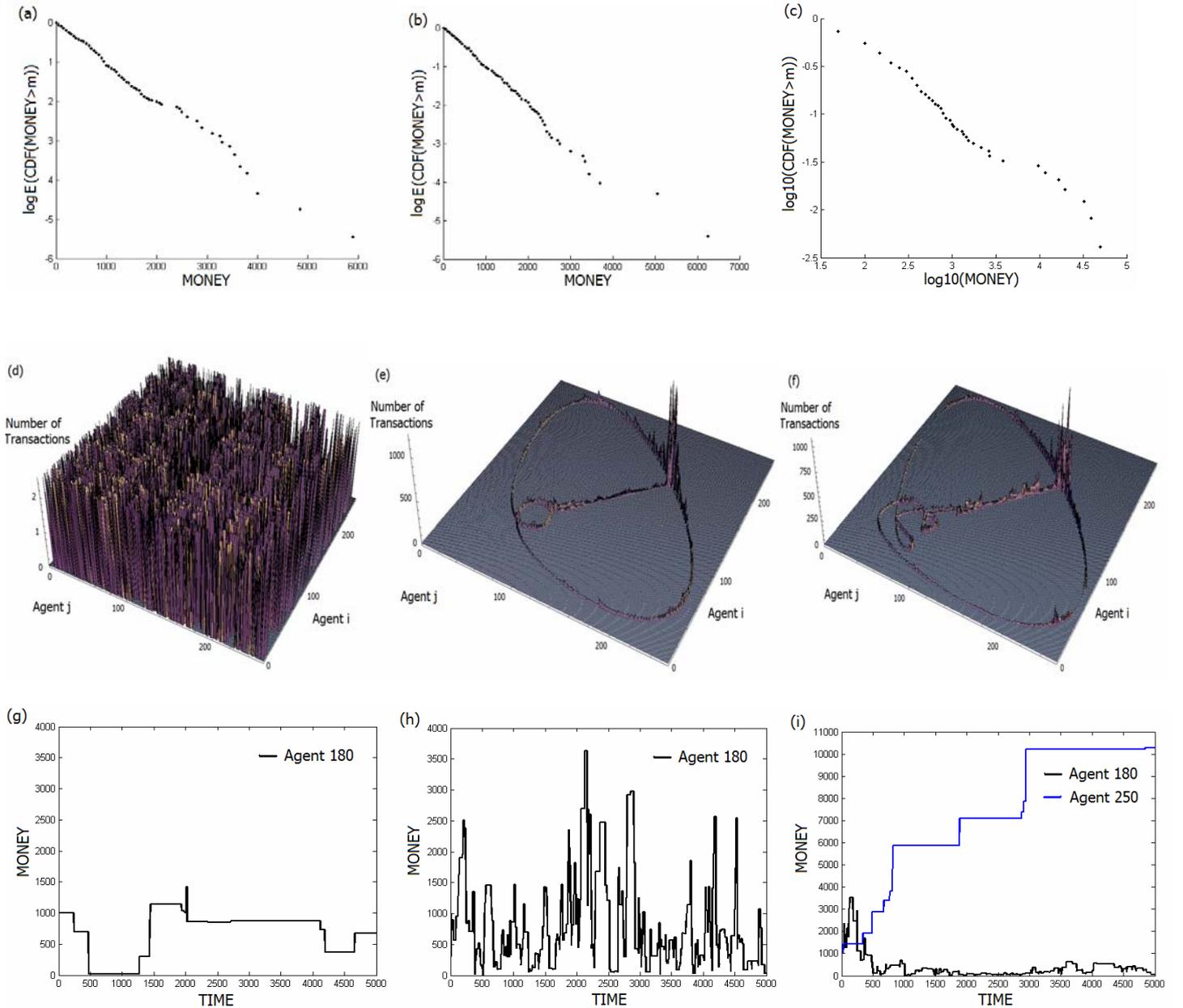

**Figure 3. Cumulative Distribution Function (CDF) of final wealth with different generator parameters, (a) P=3, (b) $\lambda_A=\lambda_B=1.05032$ (c) $\lambda_A=1.05032$ $\lambda_B = 1.08492$. Network of interactions as a histogram of interactions of agents (i,j) with different generator parameters, (d) P=3, (e) $\lambda_A=\lambda_B=1.05032$ (f) $\lambda_A=1.05032$ $\lambda_B = 1.08492$. Evolution in time of agents' money with different generator parameters, (a) P=3, (b) $\lambda_A=\lambda_B=1.05032$ (c) $\lambda_A=1.05032$ $\lambda_B = 1.08492$.**

These results illustrate how the asymptotic distribution of wealth follows the exponential profile in the two first scenarios: selection of agents with poor random statistical quality and chaotic with a symmetric attractor. This can be observed in Fig.3 (a) and (b), where the natural logarithmic plot of the Cumulative Distribution Function (CDF) shows the exponential profile of these two distributions.

The network of agent interactions in these two scenarios is quite different; in the first one, it follows a pseudo-random pattern. See Fig.3 (d), where almost every pair *(i,j)* has registered a number transactions. Although it is said to be poorly random, this scenario produces quite uniformity in the interactions of the community. In the symmetric chaotic case, the network of interactions is quite different for agents interact only within specific groups (see Fig.3 (e)). This can be a more realistic scenario, as individuals in real life use to interact within specific groups. As in this scenario, real life individuals interact mainly with their neighbours and several distant persons.

In Fig.3 (d) and (e), it also can be observed that in the random-like or the symmetric attractor scenarios, the network of interactions shows a symmetric pattern of individual interactions in the **x**-**y** axis. This gives to every agent the same opportunities to be selected as agent *i* (looser) or *j* (winner). It resembles a society where all individuals have the same opportunities. This is the reason why the evolution of agents' money looks like random, see Fig.3 (g) and (h). Becoming rich or poor in the end, is just a question of having profitable transactions or being in the right place at the right time. On a global perspective, these "uniform" or "symmetric" scenarios show that there are no sources or sinks of wealth in the community and an equilibrium state with the exponential distribution of wealth is obtained.

In contrast, the third scenario, where the selection of agents is done with a chaotic asymmetric attractor, produces a power law distribution of wealth. This can be observed in Fig.3 (c) where representation of the CDF in a double logarithmic plot shows the power law profile. In Fig. 3 (f), it can be seen how this scenario is able to select chaotic pairs of agents where some of them are selected preferably as winners (agents *j*) and never as losers (agents *i*). It resembles a society where some individuals belong to specific circles of higher economic power.

This scenario produces a non-equilibrium situation where some agents extract money from the community and make the rest poorer as time evolves. This can be clearly observed in Fig.3 (i), where agent 250 increases its income, extracting progressively money from the rest of the community. In this scenario other agents, as number 180, no longer obtain as much money as in the previous ones.

This may give some insight of what the causes at a micro level might be, that originate unfair distributions of money in society. Ultimately, the chaotic model makes obvious the inherent instability of the asymmetric scenarios, where sinks of wealth appear in the market and doom it to complete inequality.

## 1. CONCLUSIONS

Agent-based models have provided Econophysics (and other physics-like disciplines) with a powerful and authoritative simulation tool. However, the inherent complexity of real systems persistently demands for new and improved approaches in these models.

This work contributes to these demands proposing the use chaotic number generators in agent-based models used in Econophysics. The introduction of chaos in these models is based on two considerations. One is that real Economy shows some kind of a chaotic character. The other one is that chaotic pseudo-random generators offer very simple and flexible computerized engines, able to reproduce a richer number of simulation scenarios.

This approach is fully described and its implementation is explained in detail in this paper. To do that, an example in the field of economic (ideal) gas-like models is presented. The example proposes a chaotic gas-like model and it exhibits the power and high flexibility of chaotic based number generators, as simulations tools in Econophysics.

The chaotic gas like model implements a chaotic selection of agents in a trading community of agents. It has been said that gas-like models are not able to produce the power-law distribution of wealth without considering mechanisms of savings or investments [8]. With the chaotic gas like model, it is shown how chaotic based number generators can provide easily controllable scenarios and produce transitions such as the one of exponential to power-law wealth distributions.